# Species Identification and Profiling of Complex Microbial Communities Using Shotgun Illumina Sequencing of 16S rRNA Amplicon Sequences

Swee Hoe Ong[1*], Vinutha Uppoor Kukkillaya[1*], Andreas Wilm[1*], Christophe Lay[2], Eliza Xin Pei Ho[1], Louie Low[1], Martin Lloyd Hibberd[1], Niranjan Nagarajan[1#]

1. Genome Institute of Singapore, Genome #02-01, 60 Biopolis Street, Singapore.
2. Danone Research Centre for Specialised Nutrition, #09-01/02, 11 Biopolis Way, Singapore.

[*] *Contributed Equally*
[#] *Corresponding Author* [e-mail: nagarajann@gis.a-star.edu.sg]

## Abstract

The high throughput and cost-effectiveness afforded by short-read sequencing technologies, in principle, enable researchers to perform 16S rRNA profiling of complex microbial communities at unprecedented depth and resolution. Existing Illumina sequencing protocols are, however, limited by the fraction of the 16S rRNA gene that is interrogated and therefore limit the resolution and quality of the profiling. To address this, we present the design of a novel protocol for shotgun Illumina sequencing of the bacterial 16S rRNA gene, optimized to amplify more than 90% of sequences in the Greengenes database and with the ability to distinguish nearly twice as many species-level OTUs compared to existing protocols. Using several *in silico* and experimental datasets, we demonstrate that despite the presence of multiple variable and conserved regions, the resulting shotgun sequences can be used to accurately quantify the constituents of complex microbial communities. The reconstruction of a significant fraction of the 16S rRNA gene also enabled high precision (>90%) in species-level identification thereby opening up potential application of this approach for clinical microbial characterization.

# Introduction

The use of high-throughput short-read sequencing of the 16S rRNA amplicon for the profiling of microbial communities has become an increasingly attractive option for researchers due to its cost-effectiveness, and this has been further aided by the capability to do multiplexed sequencing [1,2]. However, this approach is limited by its perceived lack of precision in characterizing a microbial community and the presence of amplification biases for various variable regions of the 16S rRNA gene [3,4,5,6]. In particular, all published Illumina protocols have been restricted by an approach based on end-sequencing of specific short variable regions [7,8,9,10,11,12,13], due in part to the fragment-size limitations for paired-end Illumina sequencing, but also due to the bioinformatics challenge in a combined analysis of short-reads from different variable regions.

Correspondingly, despite being substantially more costly, 454 pyrosequencing of 16S rRNA amplicon sequences is still a popular approach in the scientific community [14] as the longer reads can provide more reliable and specific matches and enable easier analysis (though recent studies suggest that longer read lengths occasionally may not provide more information [15,16]). This advantage is in part offset by the presence of homopolymer errors and the lower read counts that impact the identification of rare and novel taxa. Furthermore, 454 pyrosequencing is currently prohibitively expensive for clinical microbiome studies that often involve hundreds of samples and multiple time points. Therefore, improved application of relatively inexpensive short-read sequencing platforms is a critical need.

In this study, we report a shotgun short-read sequencing approach (developed on, but not specific to the Illumina platform) for reconstructing 16S rRNA amplicon sequences that we demonstrate is a) less biased and tuned to capture a greater fraction of 16S rRNA gene sequences and b) provides accurate assignment (precision > 90%) at deeper taxonomic levels using the sequences. While the advantage of shotgun sequencing of a significant fraction of the 16S rRNA gene is almost self-evident, it is not clear if the resulting short read data can be assembled reliably and used effectively. In this work, we demonstrate using several *in silico* and experimental datasets that the resulting shotgun short reads can be precisely re-assembled into amplicon sequences for characterizing the constituents of complex microbial communities. Significantly, the ability to accurately reconstruct sequences enables, to our knowledge, the first reported approach for accurate species-level identification based on the 16S rRNA gene using Illumina sequencing. This makes our approach valuable for clinical applications and represents a step in the direction of routine microbial diagnostics based on high-throughput sequencing.

## Materials and Methods

*Sample collection*

Stool samples were collected from a healthy 34-year-old adult and a healthy 2-year-old infant. Throat swab sample SW18 is from an adult with macular degeneration whereas 50658 is from a healthy control. A 33-species artificial bacterial community of known composition, referred to here as ABC33 (for Artificial Bacterial Community; see **Supplementary Table S1** in **File S1**), was created by pooling equimolar concentrations of bacterial genomic DNA acquired from the American Type Culture Collection (ATCC), the Deutsche Sammlung von Mikroorganismen und Zellkulturen (DSMZ), and the Japan Collection of Microorganisms (JCM).

*Nucleic acid extraction from swab samples*

Swabs were broken off and placed in Lysing Matrix E tubes (MP Biomedicals). 360µl of Buffer ATL (QIAGEN) was added and samples were homogenised at 6m/s for 40 seconds on FastPrep Automated Homogenizer (MP Biomedicals). The suspension was centrifuged at full speed for 1 minute. 20µl of Proteinase K (QIAGEN) was mixed thoroughly with homogenized supernatant and incubated for 30 mins at $56^0$C. Next, 200µl of Buffer AL (QIAGEN) was added and vortexed, followed by 200µl of 96-100% ethanol.

The mixture was transferred into a DNAeasy Mini Spin column and centrifuged at ≥6000x g for one minute, and the eluate discarded. This step was done for the first time with the addition of 500µl of Buffer AW1 (QIAGEN) and repeated for a second time with 500µl of Buffer AW2 (QIAGEN). DNA elution was done using 50µl of Buffer EB (QIAGEN) and stored at $-20^0$C.

*Nucleic acid extraction from stool samples*

100-200 mg of stool sample was weighed in a microcentrifuge tube and transferred into Lysing Matrix E tube (MP Biomedicals) before adding 1.4ml of Buffer ASL (QIAGEN). Samples were homogenised twice at 4m/s for 30secs using a FastPrep Automated Homogenizer (MP Biomedicals). The suspension was next heated at 95°C for 5 mins and centrifuged at full speed for 1 min. One InhibitEX tablet was added to each sample and vortexed for 1 min. The suspension was incubated for 1 min at room temperature before centrifuging at full speed for 3 mins. 15μl of Proteinase K (QIAGEN) and 200μl Buffer AL (QIAGEN) were added to the supernatant and incubated for 10 mins at 70°C.

Next, 200μl of 96-100% ethanol was added and samples were transferred into QIAamp Spin columns (QIAGEN). The columns were centrifuged at full speed for 1 min, and the eluate discarded. This step was done for the first time with the addition of 500μl of Buffer AW1 (QIAGEN) and repeated for a second time with 500μl of Buffer AW2 (QIAGEN). DNA elution was done using 200μl of Buffer AE (QIAGEN) and stored at -20°C.

*Evaluation of PCR primer universality*

PCR primer sequences were compared against the sequences in three popular 16S rRNA databases namely Greengenes (dated May 9, 2011) [17], RDP (Release 10 Update 27) [18] and SILVA (Release 108) [19]. A perfect match (after fully accounting for ambiguous letters) between the primer sequence and a subsequence in the database entries was considered a hit.

*16S rRNA gene amplification and sequencing*

Bacterial 16S rRNA gene sequences were amplified using the primer pair 338F* (5'-ACTYCTACGGRAGGCWGC-3') and 1061R (5'-CRRCACGAGCTGACGAC-3'); see Figure 1 and related text for details. Briefly, each 50 μL of polymerase chain reaction (PCR) reaction contains 100 ng of fecal genomic DNA or 3 μL of throat swab genomic DNA respectively as template, 10μL 5X HotStar HiFidelity PCR buffer, 0.5 μM of each primer, 1 μL of HotStar HiFidelity DNA polymerase (2.5U) and 1 μL of 25 mM $MgSO_4$ (all part of the HotStar HiFidelity Polymerase Kit from QIAGEN).

PCR reactions were carried out using the respective protocols: (1) for the stool samples, an initial denaturation step performed at 95°C for 5 min followed by 30 cycles of denaturation (95°C, 30s), annealing (59°C, 30s) and extension (72°C, 1 min), and a final elongation of 6 min at 72°C;

(2) for ABC33 and throat swabs, the parameters were the same as above but we used 35 PCR cycles.

PCR products between 700 and 1,000 bases in size were then purified using QIAquick PCR Purification Kit (QIAGEN) and quantified using NanoDrop (Thermo Fisher Scientific). Purified amplicons were then sheared in a controlled manner to fragments with an average length of 180 bases using Adaptive Focused Acoustics$^{TM}$ (Covaris). DNA sequencing libraries were constructed from the fragments using NEBNExt® DNA Sample Preparation Reagents (New England Biolabs) according to the manufacturer's protocol.

DNA sequencing libraries were labeled with different multiplex indexing barcodes using the Multiplexing Sample Preparation Oligonucleotide Kit from Illumina. Finally, multiplexed paired-end sequencing (2×75bp reads) of the sheared fragments was done using an Illumina Genome Analyzer IIx.

*Pre-processing of sequencing datasets*

Image analysis and base calling were done on the Genome Analyzer IIx using CASAVA 1.7. After de-multiplexing of data and removal of reads that failed Illumina's purity/chastity filter (PF=0), reads were converted to FASTQ format. Reads were then filtered and trimmed by removing trailing bases with quality scores lower or equal to 2, followed by removal of read pairs containing reads shorter than 60 bases [20].

*Resolution of sequencing approaches*

In order to provide a theoretical measure for the resolution of various primer combinations and sequencing approaches, corresponding regions were extracted (the whole amplicon for shotgun sequencing) from 16S rRNA sequences in the Greengenes database (dated May 9, 2011; current_prokMSA_unaligned.fasta; **Table 1**) and clustered using UCLUST [21] (version 2.0.591; parameters: --optimal) at the species (97% identity) and genus level (95% identity). Clusters were also assessed for purity i.e. the percentage of clusters that do not have discordant species or genus level taxonomy assignments, based on the taxonomy assignments provided in Greengenes (current_GREENGENES_gg16S_unaligned.fasta: OTU ids were used at the species level).

*Generation of simulated datasets*

Three simulated datasets were generated based on three community composition profiles ("Oral", "Gut" and "Complex") using the metagenomic simulator MetaSim [22]. The composition of the "Gut" community was determined based on 2,062 16S rRNA gene sequences (DDBJ/ EMBL/GenBank accession numbers DQ325545 to DQ327606) reported by Gill *et al*. [23]. Sequences were searched using BLAST against a pruned version of the Greengenes database which only contains sequences for which taxonomic information is specified down to the species-level, and the top BLAST hit returned (E-value < 1e-4; all sequences had hits) was collected to generate a composition profile. The composition of the "Oral" microbiome was based on 14,115 16S rRNA gene sequences found in human saliva samples reported by Nasidze *et al*. [24] and determined in a similar fashion (Mihai Pop, personal communication). The composition of a "Complex" community was obtained from Turnbaugh *et al*. [25] (**Supplementary Table S3** in **File S1**, "Uneven 1") and contains 67 taxa of vastly varying abundances ranging from 0.000103% to 10.3%.

The simulation options for MetaSim were set to mimic features of the sequencing dataset for ABC33. For each community a total of 3.3 million paired-end reads of length 75 bases with an insert size of 160 and a standard deviation of 40 were simulated. The sequencing error profile for MetaSim was derived from the base-pair quality scores averaged per position of the ABC33 dataset. Quality values used for generating the error profile were then uniformly applied to all simulated sequences to obtain valid FASTQ files.

*Reconstruction of 16S rRNA amplicon sequences*

The expectation-maximization based assembly program EMIRGE [20], originally designed for whole genome datasets, was adapted to help reconstruct the amplicon sequences from the short-read datasets. Specifically, to reduce resource usage and runtime, the analysis was limited to the top (in terms of average quality) 100,000 reads, where the results were confirmed to be robust to sampling (**Supplementary Table S2** in **File S1**). EMIRGE (GIT version 98787b5) was run with parameters set to match known read and insert lengths and sequences with relative abundance below 0.1% were filtered out (except when stated otherwise).

*Classification of amplicon sequences*

Sequences reconstructed by EMIRGE were trimmed to the primer amplified regions and searched using BLAST against the complete Greengenes database (dated May 9, 2011; *current_GREENGENES_gg16S_unaligned.fasta*). BLAST hits were sorted in consecutive order by lowest E-value, highest bit-score, highest percent identity and longest alignment length, and only the top hit according to these sorting criteria was used for classification. Hits below predefined percent identity (97% at the species level, 95% at the genus level and 80% at the phylum level) were not considered for classification purposes and dropped. Note that the dropped hits are either sequences incorrectly reconstructed by EMIRGE or novel sequences that do not have similar enough sequences in the Greengenes database.

Classification results from EMIRGE (and modQIIME and RTAX, see below) were evaluated in terms of precision (=TP/(TP+FP)) and recall (=TP/(TP+FN)). A hit was considered a true positive (TP) if it matched the classification (at the appropriate level, species or genus) of a member of the simulated community, and was otherwise marked a false positive (FP). Members of the simulated community with relative abundance above the appropriate threshold (typically 0.1%, except when stated otherwise) that did not have a true positive hit were marked as false negatives (FN).

*Characterization of community composition*

Sequences reconstructed by EMIRGE are also assigned abundance estimates by the program and this enabled us to use EMIRGE results to directly characterize community composition at various taxonomic levels. As an alternative to EMIRGE, we evaluated the generic 16S rRNA analysis pipeline QIIME version 1.3.0 [26] for its ability to provide higher recall rates and thus be more sensitive in detecting constituents of a community. Specifically, we assigned operational taxonomic units (OTUs) to the reads by using QIIME's "OTU reference" option (pick_otus:otu_picking_method=uclust_ref, pick_otus:similarity=0.97) with a pre-clustered Greengenes database (*gg_97_otus_4feb2011.fasta),* with the reverse strand matches option enabled (pick_otus:enable_rev_strand_match=True) and sequences with relative abundance below 0.1% filtered out (the false positive rate was found to increase quickly at lower thresholds). To extend QIIME to handle paired-read data, the pipeline was run separately for each of the two read files and the results were merged with a filtering step that accepts a read classification only if both ends of a read were mapped to the same OTU. Note that this approach (modQIIME) has greater precision when compared to the single read version (**Supplementary Table S3** in **File S1**) and a more sophisticated alternative called RTAX [16] is now available as part of the QIIME package.

## Results

*Tuned selection of 16S rRNA amplicons*

As a result of fragment size limitations, existing Illumina end-sequencing protocols (with reads from the ends of an amplified region) for the 16S rRNA gene have been limited in the choice of primer combinations that could be explored. Our extension to a shotgun sequencing approach enabled us a wider choice of primer combinations and the opportunity to tune it better for a desired optimization criterion. In particular, we used an *in silico* assessment to identify primers likely to minimize the number of species whose 16S rRNA genes are not amplified.

Our results clearly highlight that the three top-performing primers at the 5' end are 338F*, 533R* and 341F, whereas at the 3' end, 1061R is the standout best-performing primer (**Figure 1A** and **Supplementary Table S4** in **File S1**). Assessment of all primer combinations further emphasized the advantage of these three combinations – 338F*/1061R, 533R*/1061R, 341F/1061R – with each primer pair capable of amplifying more than 90% of the sequences in the Greengenes database (**Figure 1B**; note that a similar analysis can also be done using the package PrimerProspector [27]). This is when only perfect matches are considered as hits and therefore an even higher percentage is likely to be amplified in practice. As a longer amplicon implicitly contains more information about the corresponding 16S rRNA gene segment (see below), we selected the combination 338F*/1061R (covering 92% of sequences in the Greengenes database), which amplifies the region covering V3 to V6 of the 16S rRNA gene, for the rest of our analysis. The primer pair 338F*/1061R was also evaluated using NCBI BLAST and the UCSC In-Silico PCR software against several human genome assemblies (hg16, hg17, hg18 and hg19) to confirm that no amplification artifacts are expected – a consideration which is of relevance in the clinical context.

As shown in **Table 1**, shotgun sequencing approaches have a substantial advantage over end-sequencing protocols, having on average twice as many species-level OTUs that can be identified, in principle. Among end-sequencing protocols, sequencing the 3'-end of the V3 regions provides the greatest resolution, though the 5'-end is substantially less informative. In contrast, both ends of the V6 region can resolve more than 24,000 OTUs, possibly explaining the popularity of this choice in published studies (**Supplementary Table S5** in **File S1**). As expected, clusters produced from whole-amplicon sequences also had significantly higher purity (**Table 1**). A similar pattern was observed at the genus level, although the best end-sequencing protocol (sequencing the 3'-end of the V3 region) is comparable in resolution to shotgun sequencing approaches. Cluster purity was in general higher at the genus level and whole-amplicon sequences were uniformly better than end-sequencing approaches. Among shotgun protocols, the choice of 338F*/1061R is marginally better in resolution than 533R*/1061R and 341F/1061R at the species level but is a clearer winner at the genus level. Overall, 338F*/1061R performed the best under all metrics (**Figure 1** and **Table 1**) and was the primer pair of choice in this study.

*Precise reconstruction of the 16S rRNA gene from shotgun sequences*

While shotgun sequencing of the V3 to V6 region of the 16S rRNA gene has the potential to more completely capture microbial OTUs and with greater resolution (Table 1), accurate reconstruction of the region from short reads is a potential challenge. In our analysis, we used several *in silico* datasets ("Oral", "Gut" and "Complex") as well as real sequencing data from an artificial bacterial community (ABC33) to assess the capability of the 16S rRNA gene sequence assembler EMIRGE.

Similar to the results in the original paper [20] based on whole-genome shotgun sequencing data, EMIRGE was able to reconstruct sequences with precision consistently higher than 90% at the genus as well as at the species level, and even achieved perfect precision at the species level for the *in silico* "Gut" community (**Table 2**). The few false positives reported were found to match species closely related to the true positives and may have arisen due to the limitations of the "best BLAST hit" criterion we adopted for classification. To our knowledge, this is the first report of precise species-level identification using Illumina sequencing of the 16S rRNA gene.

*Precision/Recall tradeoffs using modQIIME*

Our evaluation of amplicon sequences from EMIRGE suggests that while the "reconstruction followed by classification" approach can result in high precision, recall rates, especially at the species level, may be low for some communities. This observation could be a function of the conservative reconstruction approach employed by EMIRGE. However, our naive BLAST-based classification could also be the culprit and more sophisticated algorithms could potentially lead to higher recall rates.

Trading off precision in order to achieve a higher recall, we explored a modified clustering-based approach that directly classifies reads without an intermediate reconstruction step (modQIIME). Our results (**Table 2**) suggest that at the genus level, we can indeed do this tradeoff and obtain recall rates higher than 90%, with a variable loss in precision. An alternative tradeoff, typically intermediate between EMIRGE and modQIIME, is also possible using RTAX [16] which has recently become available as part of the QIIME package (**Table 2**). Species-level classifications however continued to have modest recall rates using clustering-based approaches, suggesting that EMIRGE would be more appropriate for this task. Note that species-level recall rates for all approaches and, in particular for EMIRGE, was significantly lower for the "Complex" and ABC33 datasets, highlighting the challenge of species-level identification when many closely-related species are present in a community (**Table 2**). In terms of diversity metrics, all three approaches mostly over-estimated the diversity of the samples (**Supplementary Table S6** in **File S1**), with RTAX and EMIRGE typically being the closest to the true answer. Both EMIRGE and modQIIME were moderately compute and memory intensive (typically taking a few hours and <11 hours with 4 CPUs and <25Gb of RAM) while RTAX required several days to analyze a dataset in the worst case.

*Concordance of microbial community structure*

A strong advantage of deep sequencing of the 16S rRNA gene on the Illumina platform is the potential to accurately quantify abundances for even rare members of a microbial community. Our analysis of the *in silico* datasets suggests that the abundances estimated from the reconstructed sequences were indeed quite accurate even at the species level (**Figure 2A**) and with correlation coefficients greater than 0.95 for EMIRGE on all datasets. The clustering approaches (modQIIME and RTAX), generally have poor correlation coefficients at the species level (-0.2 to 0.7), but have modest results at the genus level (correlation coefficient > 0.7).

Analysis of sequencing datasets from the throat swab and stool samples using EMIRGE and modQIIME showed a broad agreement in their results and with what is known about these microbial communities through Sanger and 454 sequencing (**Figure 2B**). For example, the stool microbiota was dominated by Bacteroidetes and Firmicutes, followed by Actinobacteria and Tenericutes, in agreement with previous Sanger [28,29] and 454 [30] sequencing surveys. The most notable compositional difference between the infant and adult stool samples is the difference in their Bacteroidetes:Firmicutes ratio [31,32,33], with a lower percentage of Firmicutes observed in the 2-year-old infant compared to the 34-year-old adult.

For the throat swab samples, the five most abundant phyla detected were Actinobacteria, Bacteroidetes, Firmicutes, Proteobacteria and Fusobacteria, which is in agreement with 454 sequencing results reported by Jakobson *et al*. [34] and with 16S rRNA gene microarray by Lemon *et al.* [35]. The high variability in composition between samples has also been noted before from saliva [24], and in particular in this study, throat swab sample SW18 has a much higher abundance of Actinobacteria and lower abundance of Fusobacteria compared to sample 50658. Interestingly, our analysis revealed a significant proportion of sequences that could not be annotated at the species-level (>15% in terms of relative abundance for SW18; though they can be classified at the phylum-level - **Figure 2B**), highlighting the strength of our approach for studying novel constituents of a microbial community.

Analysis at the species level identified 38 members in the two stool samples and 44 members in the two throat swabs we sequenced, with several members detected at as low as 0.01% abundance (**Supplementary Tables S7** and **S8** in **File S1**). Interestingly, at the species level the infant and adult stool samples have few species in common whereas the throat samples share most of their abundant members. As a sanity check, we also confirmed that a majority of the reported species are common constituents of the gut and oral microbiota (**Supplementary Table S7** in **File S1**).

# Discussion

The novel shotgun 16S rRNA Illumina sequencing protocol presented here has clear theoretical advantages, with a primer pair optimized to amplify a longer stretch of the 16S rRNA gene as well as more sequences (92% of the Greengenes database) and selected to have high resolution at both the genus and species level. Our empirical results further highlight its utility for precise (>90% at the species level) and high-resolution microbiome profiling, though additional benchmarking using long-read sequencing datasets would be ideal. Taken together, we believe this makes for a good case for wide usage of this protocol (especially when species-level classification is desired) on the Illumina platform. While the read lengths analyzed here were around 75bp, longer reads (up to 150bp) can currently be generated on an Illumina HiSeq at a greater cost and with higher sequencing error rates (even longer reads of up to 250bp can be generated for a significantly higher cost on the MiSeq). These longer reads should allow for more precise reconstruction and analysis and as read lengths approach the typical amplicon length (this is already possible on a PacBio *RS* sequencer but at a much greater cost), computational analysis of the resulting sequences will get simplified.

With recent improvements in sequencing throughput, using deep DNA sequencing as a pathogen screening tool is an attractive idea but its utility is limited by contamination from non-microbial and host DNA. The use of 16S rRNA amplicon sequencing can address this drawback but it comes with the cost of amplification biases. Our results for the sequencing and analysis protocol presented here suggests that with a careful choice of primers, the biases can be minimized, and that microbial constituents of a sample can be precisely quantified at the species level using as few as 100,000 reads. With improved automation of library-preparation and multiplexing steps, this approach will be cost and time effective for future clinical microbiome studies with hundreds of samples and multiple time points. A recent example of such a study is one that looked at the association of gut microbiota with type 2 diabetes and uncovered potential biomarkers [36]. A 16S rRNA-based approach such as the one described here would be more cost effective when similar studies are conducted for microbiota of body sites where host DNA contamination can be significant (e.g. oral and skin).

The principal approaches used for short-read sequence analysis in this study (EMIRGE and modQIIME) were moderately compute and memory intensive (requiring large clusters if hundreds of samples need to be analyzed) and had modest recall rates. Improved algorithms for data analysis could potentially enable better tradeoffs between compute resources, sequencing depth and sensitivity for reliable detection of rare species in a microbial community.

## Acknowledgements

We thank the Genome Technology & Biology group at GIS for sequencing and the Scientific & Research Computing group for sequence analysis support. We would also like to thank the anonymous reviewers for valuable comments and suggestions to improve the manuscript.

## Availability

Simulated datasets and community profiles can be found at http://collaborations.gis.a-star.edu.sg/~shotgun_16S_sequencing/. The post-processor script for modQIIME can be found at https://github.com/CSB5/16s-arxiv-1210.3464 (version 1). The five human sample datasets (adult gut, infant gut, throat SW18, throat 50658 and ABC33) can be accessed from NCBI Sequence Read Archive (SRA) via accession numbers SRX148649-148652.

**Figure Legends**

**Figure 1. *In silico* evaluation of 16S rRNA PCR primers.** A) Percentage of sequences matching individual primers, with the top two primers highlighted in boxes. B) Percentage of sequences amplifiable by various primer pairs (338F*/1061R is the best pair). Percentage of matched sequences is measured against the Greengenes 16S rRNA sequence database. See **Supplementary Table S4** in **File S1** for primer sequences and results measured against the RDP and SILVA databases. Primer numbering is based on the *E. coli* system of nomenclature as in Brosius *et al*. [37] and for simplicity the same name (say 784F) is used for both forward and reverse primers at a given position.

**Figure 2. Community composition based on 16S rRNA sequence reconstruction using EMIRGE.** A) Correlation between known and estimated relative abundances of predicted species on three *in silico* datasets. A log-scaled version of this plot can be seen in **Supplementary Figure S1** in **File S1**. B) Composition at the phylum level for the throat swab and stool sequencing datasets.

## Tables
### Table 1

| Sequencing Approach | Reads From | Species-level OTUs | Genus-level OTUs |
|---|---|---|---|
| End sequencing | | | |
| V3 (338F*/533R*) | 5'-end | 7,388 (76%) | 4,526 (83%) |
| | 3'-end | 35,763 (92%) | 27,699 (**97%**) |
| V4 (533R*/805R) | 5'-end | 10,971 (83%) | 6,671 (88%) |
| | 3'-end | 15,000 (87%) | 9,993 (92%) |
| V5 (805R/907F) | 5'-end | 23,301 (91%) | 17,138 (96%) |
| | 3'-end | 10,501 (83%) | 6,746 (89%) |
| V6 (907F/1061R) | 5'-end | 3,701 (73%) | 2,221 (77%) |
| | 3'-end | **39,886 (92%)** | 31,285 (96%) |
| Shotgun sequencing | | | |
| V3-V6 (338F/1061R) | Whole amplicon | 59,378 (97%) | 34,869 (99%) |
| V3-V6 (338F*/1061R) | Whole amplicon | **61,298 (97%)** | **36,361 (99%)** |
| V3-V6 (341F/1061R) | Whole amplicon | 59,272 (97%) | 35,109 (99%) |
| V4-V6 (533R*/1061R) | Whole amplicon | 59,436 (97%) | 35,161 (99%) |

**Table 1. Species- and genus-level resolution of various sequencing approaches.** Resolution was measured by the number of OTUs/clusters produced using UCLUST [21] at the species (97% identity) and genus level (95% identity) for 16S rRNA sequences in the Greengenes database, based on various end-sequencing (76 bases in length from either the 5' or 3' end) and shotgun-sequencing approaches and primer combinations. A higher OTU/cluster number indicates a theoretical higher level of resolution for taxonomic classification. The numbers in parenthesis provide the purity of clusters as measured by the percentage of clusters with homogenous taxonomy assignments in Greengenes. Entries with the highest resolution and/or purity for each sequencing approach are marked in bold. The primer sequences can be found in **Supplementary Table S4** in **File S1**.

**Table 2**

| Method | Genus-level recall (%) | Genus-level precision (%) | Species-level recall (%) | Species-level precision (%) |
|---|---|---|---|---|
| *"Oral"* | | | | |
| EMIRGE (33%) | 88 | 90 | 66 | 96 |
| modQIIME (93%) | 97 | 63 | 66 | 51 |
| RTAX (95%) | 88 | 88 | 61 | 68 |
| *"Gut"* | | | | |
| EMIRGE (30%) | 84 | 95 | 69 | 100 |
| modQIIME (92%) | 92 | 82 | 71 | 94 |
| RTAX (92%) | 88 | 76 | 82 | 77 |
| *"Complex"* | | | | |
| EMIRGE (13%) | 64 | 100 | 32 | 86 |
| modQIIME (78%) | 100 | 55 | 59 | 59 |
| RTAX (86%) | 76 | 53 | 49 | 38 |
| *ABC33* | | | | |
| EMIRGE (60%) | 83 | 94 | 39 | 93 |
| modQIIME (95%) | 94 | 85 | 48 | 70 |
| RTAX (96%) | 100 | 90 | 52 | 61 |

**Table 2. Evaluation of EMIRGE, modQIIME and RTAX on different datasets.** Precision and recall rates for the "Oral", "Gut", "Complex" and ABC33 datasets using EMIRGE, modQIIME and RTAX at a 0.1% relative abundance threshold. The percentage of sequences/OTUs removed because of the abundance threshold is given in parentheses for each method.

A

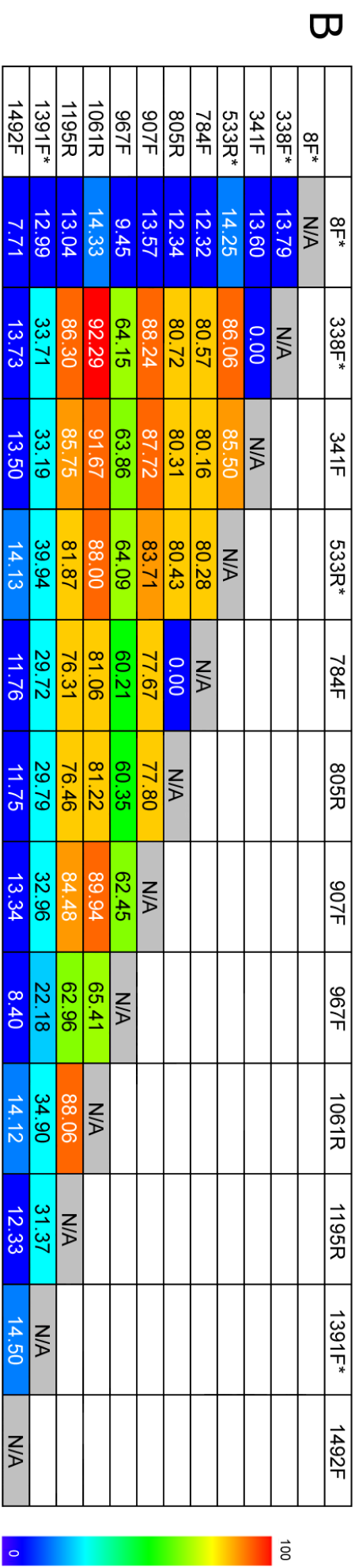

B

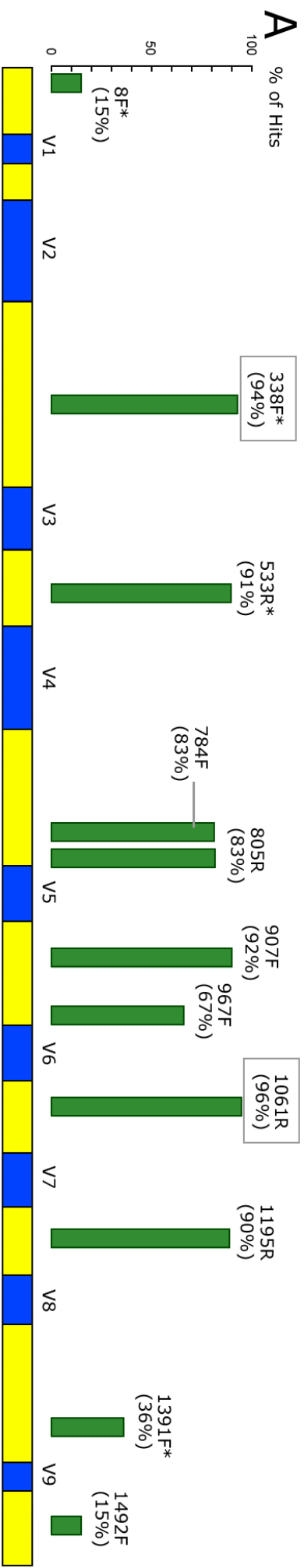

| | 8F* | 338F* | 341F | 533R* | 784F | 805R | 907F | 967F | 1061R | 1195R | 1391F* | 1492F |
|---|---|---|---|---|---|---|---|---|---|---|---|---|
| 8F* | N/A | | | | | | | | | | | |
| 338F* | 13.79 | N/A | | | | | | | | | | |
| 341F | 13.60 | 0.00 | N/A | | | | | | | | | |
| 533R* | 14.25 | 86.06 | 85.50 | N/A | | | | | | | | |
| 784F | 12.32 | 80.57 | 80.16 | 80.28 | N/A | | | | | | | |
| 805R | 12.34 | 80.72 | 80.31 | 80.43 | 0.00 | N/A | | | | | | |
| 907F | 13.57 | 88.24 | 87.72 | 83.71 | 77.67 | 77.80 | N/A | | | | | |
| 967F | 9.45 | 64.15 | 63.86 | 64.09 | 60.21 | 60.35 | 62.45 | N/A | | | | |
| 1061R | 14.33 | 92.29 | 91.67 | 88.00 | 81.06 | 81.22 | 89.94 | 65.41 | N/A | | | |
| 1195R | 13.04 | 86.30 | 85.75 | 81.87 | 76.31 | 76.46 | 84.48 | 62.96 | 88.06 | N/A | | |
| 1391F* | 12.99 | 33.71 | 33.19 | 39.94 | 29.72 | 29.79 | 32.96 | 22.18 | 34.90 | 31.37 | N/A | |
| 1492F | 7.71 | 13.73 | 13.50 | 14.13 | 11.76 | 11.75 | 13.34 | 8.40 | 14.12 | 12.33 | 14.50 | N/A |

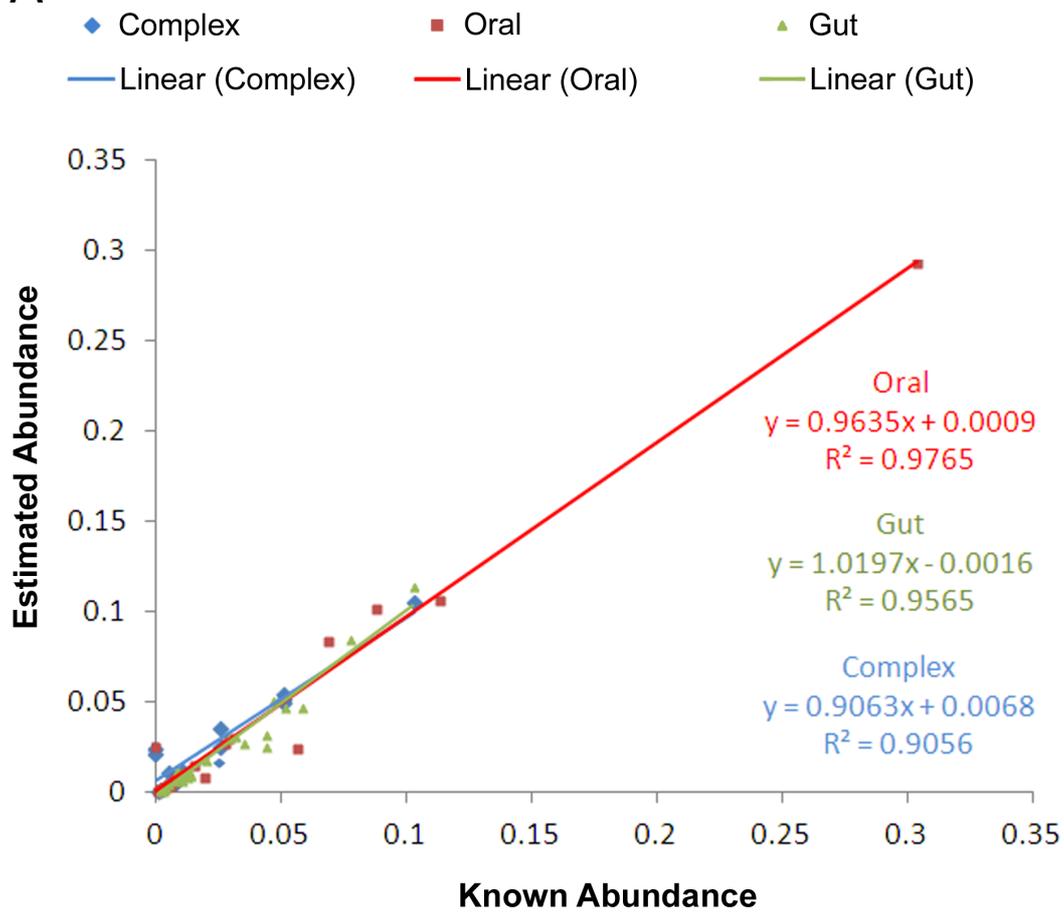

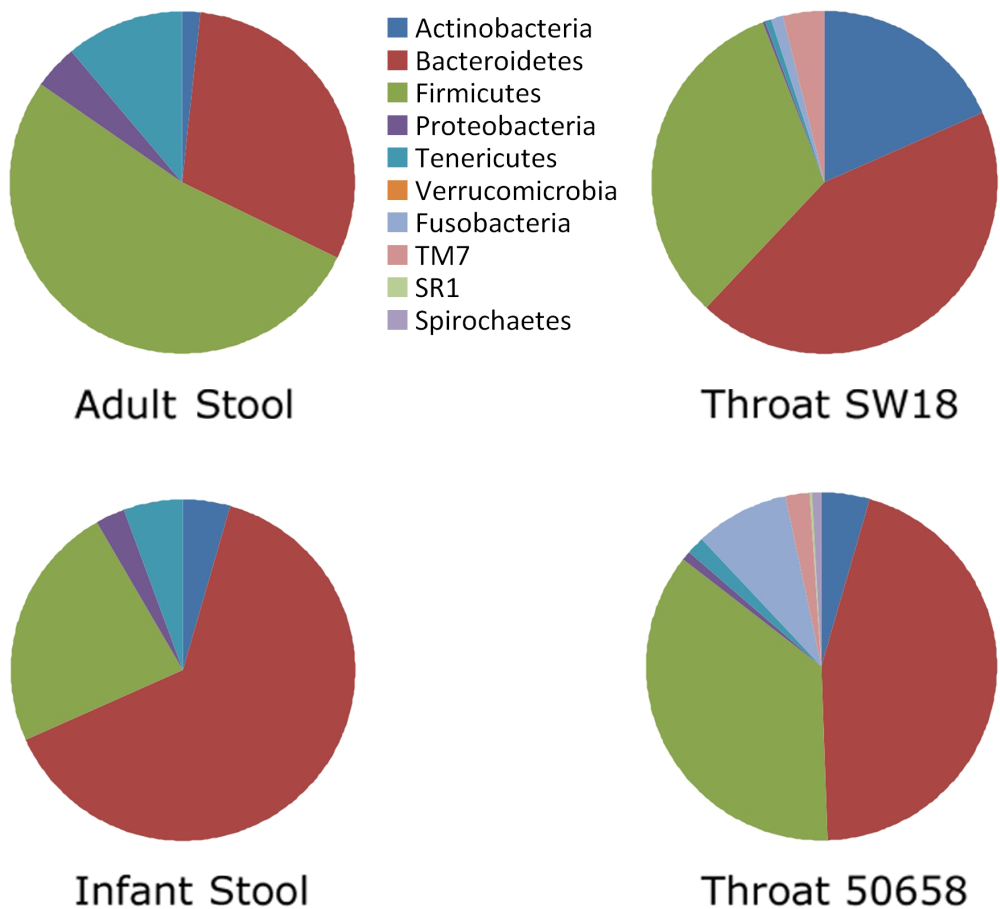

# Supplementary Information

**Supplementary Figure S1.** Log-scaled version of the plot in Figure 2A.

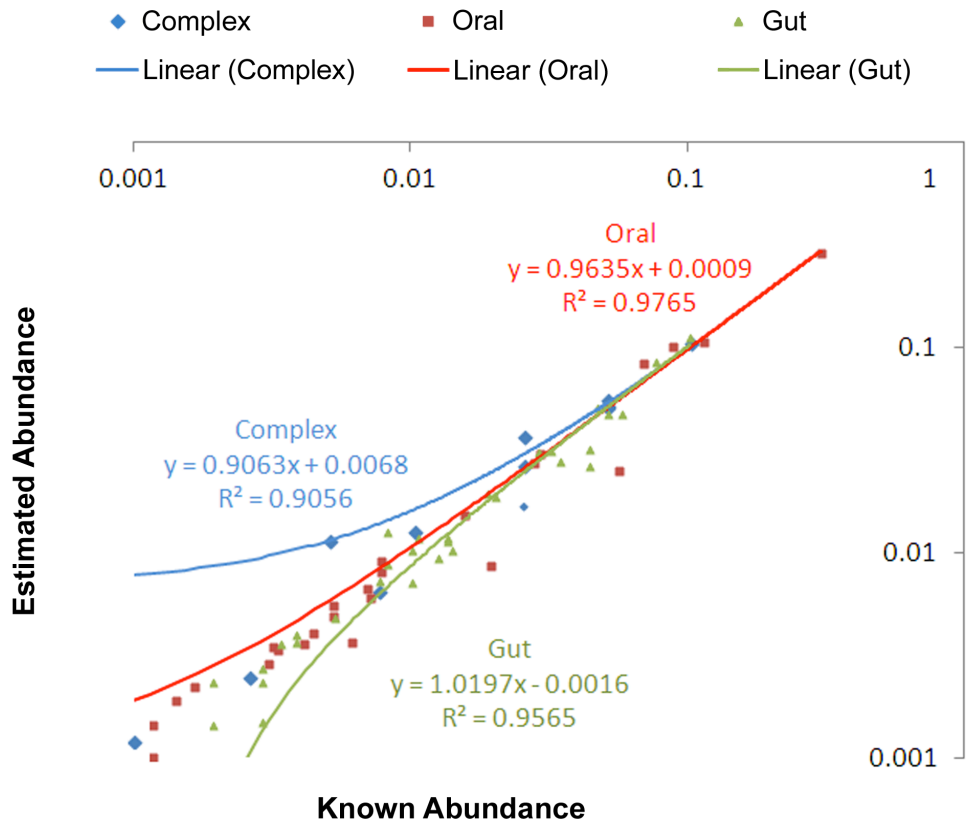

**Supplementary Table S1. Constituent bacterial species of ABC33 in alphabetical order.**

*Bacteroides fragilis* JCM 11019T
*Bacteroides thetaiotaomicron* JCM 5827T
*Bacteroides vulgatus* JCM 5826T
*Bifidobacterium adolescentis* JCM 1275T
*Bifidobacterium catenulatum* JCM 1194T
*Bifidobacterium longum* subsp. *infantis* JCM 1222T
*Bifidobacterium longum* subsp. *longum* JCM 1217T
*Blautia coccoides* DSM 935
*Blautia hansenii* JCM 14655T
*Blautia producta* JCM 1471T
*Clostridium difficile* DSM 1296
*Clostridium leptum* DSM 753
*Clostridium perfringens* DSM 756
*Collinsella aerofaciens* JCM 10188T
*Corynebacterium kroppenstedtii* JCM 11950T
*Enterococcus faecalis* ATCC 29212
*Escherichia coli* O157
*Faecalibacterium prausnitzii* DSM 17677
*Lactobacillus acidophilus* JCM 1132T
*Lactobacillus gasseri* JCM 1131T
*Mycobacterium tuberculosis* H37Ra ATCC 25177
*Neisseria gonorrhoeae* ATCC 53423
*Neisseria lactamica* DSM 4691
*Neisseria meningitidis* DSM 10036
*Neisseria mucosa* JCM 12992T
*Neisseria perflava* ATCC 14799
*Nocardia asteroides* JCM 3384T
*Porphyromonas gingivalis* JCM 12257T
*Propionibacterium acnes* JCM 6425T
*Roseburia intestinalis* DSM 14610
*Staphylococcus aureus* ATCC 6538P
*Streptococcus oralis* JCM 12997T
*Streptococcus thermophilus* JCM 20026

**Supplementary Table S2**. **Robustness of EMIRGE results to the number of reads used.** Results shown are for the "Complex" community and at 0.1% relative abundance threshold.

| Number of Reads | Genus-level recall (%) | Genus-level precision (%) | Species-level recall (%) | Species-level precision (%) |
|---|---|---|---|---|
| 50K | 59 | 100 | 24 | 82 |
| 100K | 64 | 100 | 32 | 86 |
| 200K | 68 | 100 | 41 | 88 |
| 500K | 64 | 100 | 43 | 89 |

**Supplementary Table S3**. **Performance of QIIME vs modQIIME.** Results shown are for the "Complex" community and at 0.1% relative abundance threshold.

| Method | Genus-level recall (%) | Genus-level precision (%) | Species-level recall (%) | Species-level precision (%) |
|---|---|---|---|---|
| QIIME | 82 | 46 | 46 | 55 |
| modQIIME | 100 | 55 | 59 | 59 |

**Supplementary Table S4.** *In silico* assessment of PCR primers for the 16S rRNA gene using Greengenes, RDP and SILVA. Primers with an asterisk appended to their names are optimized versions of the respective canonical primers which have had between 1 to 3 additional positions (underlined for clarity) replaced with degenerate nucleotides to improve their matching rate. The primer numbering is based on the *E. coli* system of nomenclature as in Brosius *et al*. [37].

|     | Primer | Sequence in 5'–3' orientation | Greengenes | RDP | SILVA | Combined |
| --- | --- | --- | --- | --- | --- | --- |
|     | Size of data set (no. of sequences) | | 1,011,632 | 1,727,996 | 618,442 | 3,358,070 |
| 1.  | 8F | AGAGTTTGATCCTGGCTCAG | 125,329 (12%) | 159,533 (9%) | 129,212 (21%) | 414,074 (12%) |
| 2.  | 8F* | AGAGTTTGATCMTGGCTCAG | 149,371 (15%) | 201,280 (12%) | 155,132 (25%) | 505,783 (15%) |
| 3.  | 338F | ACTCCTACGGGAGGCAGC | 946,219 (94%) | 1,384,005 (80%) | 486,609 (79%) | 2,816,833 (84%) |
| 4.  | 338F* | ACTYCTACGGRAGGCWGC | 954,507 (94%) | 1,400,148 (81%) | 494,810 (80%) | 2,849,465 (85%) |
| 5.  | 341F | CCTACGGGAGGCAGCAG | 947,239 (94%) | 1,396,163 (81%) | 486,859 (79%) | 2,830,261 (84%) |
| 6.  | 533R | GTGCCAGCAGCCGCGGTAA | 910,062 (90%) | 1,342,620 (78%) | 566,000 (92%) | 2,818,682 (84%) |
| 7.  | 533R* | GTGCCAGCMGCCGCGGTAA | 918,217 (91%) | 1,347,229 (78%) | 580,549 (94%) | 2,845,995 (85%) |
| 8.  | 784F | AGGATTAGATACCCTGGTA | 837,677 (83%) | 1,142,175 (66%) | 435,323 (70%) | 2,415,175 (72%) |
| 9.  | 805R | ATTAGATACCCTGGTAGTC | 839,319 (83%) | 1,142,173 (66%) | 436,995 (71%) | 2,418,487 (72%) |
| 10. | 907F | AAACTYAAAKGAATTGACGG | 929,021 (92%) | 1,128,872 (65%) | 535,808 (87%) | 2,593,701 (77%) |
| 11. | 967F | CAACGCGAAGAACCTTACC | 674,109 (67%) | 762,533 (44%) | 319,181 (52%) | 1,755,823 (52%) |
| 12. | 1061R | CRRCACGAGCTGACGAC | 976,162 (96%) | 1,125,505 (65%) | 516,018 (83%) | 2,617,685 (78%) |
| 13. | 1195R | GAGGAAGGYGGGGAYGACGTC | 909,466 (90%) | 1,013,808 (59%) | 459,535 (74%) | 2,382,809 (71%) |
| 14. | 1391F | TGYACACACCGCCCGTC | 339,602 (34%) | 384,798 (22%) | 413,479 (67%) | 1,137,879 (33%) |
| 15. | 1391F* | TGYACWCACYGCCCYGTC | 368,947 (36%) | 411,983 (24%) | 444,247 (72%) | 1,225,177 (36%) |
| 16. | 1492F | AAGTCGTAACAAGGTA | 149,137 (15%) | 171,929 (10%) | 162,191 (26%) | 483,257 (14%) |

**Supplementary Table S5.** Sequencing and analysis protocols in published Illumina-sequencing-based 16S rRNA studies.

| Reference | V Region | Read length | Taxonomic Assignment Method |
|---|---|---|---|
| Lazarevic *et al*., 2009 | V5 | 76 bp | GAST and the RDP Classifier. |
| Hummelen *et al*., 2010 | V6 | 76 bp | UCLUST clustering and BLAST. |
| Claesson *et al*., 2010 | V1/V2, V2/V3, V3/V4, V4/V5, V5/V6, V7/V8 | 101 bp | RDP Pyrosequencing Pipeline and the RDP naïve Bayesian Classifier. |
| Zhou *et al*., 2010 | V6 | 100 bp | The program *Merger* for merging the PE reads and GAST. |
| Gloor *et al*., 2010 | V6 | 76 bp | UCLUST clustering and BLAST. |
| Caporaso *et al*., 2011 | V4 | 100 bp | QIIME's wrappers for the RDP classifier. |
| Miller *et al*., 2011 | Genomic | 76 bp | EMIRGE. |
| Bartram *et al*., 2011 | V3 | 125 bp | RDP naïve Bayesian Classifier v.2.1. |
| Degnan & Ochman, 2012 | V6 | 75/100 bp | RDP Pyrosequencing Pipeline. |

**Supplementary Table S6. Diversity metrics for EMIRGE, modQIIME and RTAX on different datasets.** The metrics reported here are Chao1 (bias corrected version; [Chao, 1987]) and Shannon Entropy (separated by a comma). Both were computed with QIIME's alpha_diversity.py script with options '-m chao1' and '-m shannon' respectively. Results from Table 2 were used to calculate the diversity metrics and the known profile was used to compute values for the "True Profile" column.

| Dataset | True Profile | EMIRGE | modQIIME | RTAX |
|---|---|---|---|---|
| *"Oral"* | 38, 3.9 | 56, 4.2 | 167, 9.3 | 45, 3.3 |
| *"Gut"* | 45, 4.7 | 85, 5.5 | 190, 9.6 | 54, 3.0 |
| *"Complex"* | 29, 4.2 | 28, 4.0 | 189, 8.6 | 58, 3.4 |
| ABC33 | 33, 5.0 | 124, 6.3 | 124, 5.2 | 67, 5.6 |

**Supplementary Table S7. Species identified by EMIRGE analysis of shotgun Illumina sequencing datasets for stool and throat swab samples.** Species reported are those with relative abundance greater than 0.005% and the analysis was done with 500K reads (see **Supplementary Table S8** for validation at this threshold on *in silico* datasets). Species found in the stool samples which are listed at http://genome.wustl.edu/genomes/list/microorganisms and species found in the throat swabs which are in the Human Oral Microbiome Database (http://www.homd.org) are in bold.

| | |
|---|---|
| Adult Stool | ***Bacteroides uniformis*** (4.4%), ***Eubacterium rectale*** (1.34%), *Bifidobacterium longum* (1.27%), ***Bacteroides eggerthii*** (0.96%), ***Ruminococcus gnavus*** (0.8%), ***Faecalibacterium prausnitzii*** (0.47%), *Parabacteroides distasonis* (0.42%), ***Bacteroides stercoris*** (0.29%), ***Ruminococcus callidus*** (0.22%), ***Collinsella aerofaciens*** (0.17%), *Haemophilus parainfluenzae* (0.15%), ***Bifidobacterium adolescentis*** (0.14%), *Bacteroides massiliensis* (0.12%), *Coprococcus catus* (0.09%), *Bacteroides fragilis* (0.05%), ***Actinomyces odontolyticus*** (0.04%), *Veillonella parvula* (0.03%), *Clostridium orbiscindens* (0.03%), *Porphyromonas gingivalis* (0.03%), ***Bacteroides ovatus*** (0.02%), ***Coprococcus eutactus*** (0.02%), ***Holdemania filiformis*** (0.01%), *Porphyromonas endodontalis* (0.01%), *Prevotella pallens* (0.01%), *Eggerthella lenta* (0.01%) |
| Infant Stool | ***Ruminococcus gnavus*** (8.3%), ***Clostridium ramosum*** (1.57%), ***Bacteroides ovatus*** (1.54%), *Clostridium innocuum* (1.47%), ***Bifidobacterium breve*** (1.01%), *Bacteroides fragilis* (0.93%), *Blautia producta* (0.71%), *Parabacteroides distasonis* (0.57%), ***Clostridium spiroforme*** (0.43%), *Clostridium orbiscindens* (0.28%), *Streptococcus salivarius* (0.21%), ***Clostridium hylemonae*** (0.18%), ***Anaerostipes caccae*** (0.1%), *Escherichia fergusonii* (0.09%), ***Bacteroides vulgates*** (0.09%), *Shigella dysenteriae* (0.09%), ***Actinomyces odontolyticus*** (0.05%), *Bifidobacterium longum* (0.02%), *Clostridium sp. MLG480* (0.01%), *Coprobacillus cateniformis* (0.01%) |
| Throat SW 18 | ***Streptococcus salivarius*** (5.8%), ***Prevotella histicola*** (5.78%), ***Actinomyces odontolyticus*** (4.1%), ***Veillonella dispar*** (3.45%), ***Granulicatella adiacens*** (1.8%), ***Prevotella pallens*** (1.65%), ***Rothia dentocariosa*** (0.72%), ***Gemella sanguinis*** (0.45%), ***Solobacterium moorei*** (0.31%), ***Veillonella parvula*** (0.1%), ***Streptococcus mutans*** (0.06%), ***Lactobacillus salivarius*** (0.05%), ***Corynebacterium matruchotii*** (0.03%), ***Staphylococcus aureus*** (0.03%), ***Bifidobacterium longum*** (0.02%), ***Porphyromonas gingivalis*** (0.01%), ***Scardovia inopinata*** (0.01%), *Streptococcus thermophilus* (0.01%) |
| Throat 50658 | ***Streptococcus salivarius*** (4.39%), ***Actinomyces odontolyticus*** (3.57%), ***Porphyromonas gingivalis*** (2.84%), ***Veillonella dispar*** (1.9%), ***Prevotella pallens*** (1.31%), ***Porphyromonas endodontalis*** (1.26%), ***Prevotella histicola*** (1.14%), ***Granulicatella adiacens*** (0.85%), ***Tannerella forsythia*** (0.71%), ***Streptococcus anginosus*** (0.46%), ***Solobacterium moorei*** (0.39%), ***Prevotella intermedia*** (0.36%), ***Prevotella tannerae*** (0.29%), ***Treponema denticola*** (0.24%), ***Haemophilus parainfluenzae*** (0.23%), ***Treponema amylovorum*** (0.14%), ***Corynebacterium matruchotii*** (0.13%), ***Abiotrophia defective*** (0.12%), ***Lactobacillus salivarius*** (0.08%), ***Rothia mucilaginosa*** (0.08%), *Prevotella nanceiensis* (0.08%), ***Veillonella parvula*** (0.06%), ***Selenomonas sputigena*** (0.06%), ***Rothia dentocariosa*** (0.05%), ***Campylobacter rectus*** (0.05%), ***Prevotella nigrescens*** (0.05%), ***Prevotella baroniae*** (0.04%), ***Treponema lecithinolyticum*** (0.03%), ***Corynebacterium durum*** (0.03%), ***Treponema socranskii*** (0.03%), ***Bulleidia extructa*** (0.02%), ***Selenomonas noxia*** (0.01%), ***Streptococcus cristatus*** (0.01%), *Streptococcus pseudopneumoniae* (0.01%), *Lactobacillus mucosae* (0.01%) |

**Supplementary Table S8. Performance of EMIRGE at a relative abundance threshold of 0.005% using 500K reads.**

| Community | Genus-level recall (%) | Genus-level precision (%) | Species-level recall (%) | Species-level precision (%) |
|---|---|---|---|---|
| "Oral" | 80 | 90 | 53 | 94 |
| "Gut" | 86 | 94 | 80 | 100 |
| "Complex" | 71 | 89 | 42 | 92 |